\documentclass[twocolumn,showpacs,preprintnumbers,amsmath,amssymb]{revtex4}

\usepackage{graphicx}
\usepackage{dcolumn}
\usepackage{bm}

\begin{document}
\bibliographystyle{prsty}

\topmargin 0.01in

\newcommand{\be}{\begin{equation}}
\newcommand{\ee}{\end{equation}}

\title{Temporal Evolution Of Step-Edge Fluctuations \\ Under Electromigration Conditions}

\author{P.J. Rous}\email{rous@umbc.edu}
\author{T.W. Bole}

\affiliation{Department of Physics, \\
University of Maryland Baltimore County, \\
1000 Hilltop Circle, \\
Baltimore, MD 21250}

\date{\today}

\begin{abstract}
The temporal evolution of step-edge fluctuations under electromigration conditions is analysed using a continuum Langevin model. If the electromigration driving force acts in the step up/down direction, and step-edge diffusion is the dominant mass-transport mechanism, we find that significant deviations from the usual $t^{1/4}$ scaling of the terrace-width correlation function occurs for a critical time $\tau$ which is dependent upon the three energy scales in the problem: $k_{B}T$, the step stiffness, $\gamma$, and the bias associated with adatom hopping under the influence of an electromigration force, $\pm \Delta U$. For ($t  <  \tau$), the correlation function evolves as a superposition of $t^{1/4}$ and $t^{3/4}$ power laws. For $t \ge \tau$ a closed form expression can be derived. This behavior is confirmed by a Monte-Carlo simulation using a discrete model of the step dynamics. It is proposed that the magnitude of the electromigration force acting upon an atom at a step-edge can by estimated by a careful analysis of the statistical properties of step-edge fluctuations on the appropriate time-scale.
\end{abstract}


\maketitle

\section{Introduction}

During the past decade, continuum models and discrete lattice simulations have been applied to understand direct imaging observations of the thermal fluctuations of step edges in which the step position is monitored as a function of time~\cite{Jeong99,Giesen01}. Of particular interest has been the study of the dynamics of the equilibration of terrace width distributions where the temporal evolution of step edge fluctuations are driven by the exchange of atoms between the step and the adjacent terrace and/or by motion of adatoms along the step edge itself~\cite{Bartelt92,Khare98,Ihle98,JeongWeeks99,Flynn02,Bartelt02}.  It is well known that the position of the step edge, as described by its temporal correlation function, has a time dependence that scales as a power law with an exponent characteristic of specific atomic processes driving the step fluctuations; $t^{\beta}$. In cases where mass transport at the step is dominated by diffusion of atoms along the step edge; $\beta= 1/4$. When mass transport  proceeds {\em via} exchange of atoms between the step edge and the terrace $\beta=1/2$. Careful experiments allow a crossover from $t^{1/4}$ to $t^{1/2}$ scaling to be observed as a function of the sample temperature~\cite{Giesen99}. Further, experimental measurement of the correlation functions have been used to determine thermodynamic properties of the steps, such as the step stiffness~\cite{Jeong99,Giesen01}.

In this paper we investigate how the scaling of the step edge fluctuations is changed by the presence of an electromigration force~\cite{Sorbello97} acting upon atoms diffusing along the step edges. The primary motivation for this study is the possibility of using measurements of these changes to obtain information about the electromigration force itself.  In conducting materials, a surface electromigration force can be generated by passing an electrical current through the sample~\cite{Schumacher93,DH, Ishida94,Ishida95,Ishida98,Ishida99,Ishida96b,Rous1,Rous99,Rous00a,Rous00b,YasNat1}. In terms of a simple discrete model, the presence of the electromigration force introduces a small bias in the diffusion of atoms at the step edge, parallel to the current (and field). By convention, this bias can be expressed as an energy difference between atoms diffusing parallel or anti-parallel to the current $\Delta U = Z^{\ast}e{\bf E} a_\perp$ where ${\bf E}$ is the electric field applied to the sample, $Z^{\ast}$ is the effective valence of adatoms and $a_\perp$ is the lattice spacing perpendicular to the step edge. A characteristic property of surface electromigration is that the electromigration bias, $\Delta U$, is several orders of magnitude smaller than the other energy scales in the problem: $\gamma a_\perp$, where $\gamma$ is the step stiffness and the energy associated with thermal fluctuations; $k_{B}T$. This suggests that thermal fluctuations will completely dominate the short-time behavior of the step fluctuations with the effect of the electromigration bias emerging only on much longer time scales. Nevertheless, such time scales (of the order of seconds) are accessible to experiment offering the possibility that the observation of step fluctuations under electromigration conditions may allow us to obtain quantitative information about the magnitude of the force itself; a quantity that, to date, has been hidden from experimental study.

This paper is organized as following. In section~2 we present a continuum theory of step edge fluctuations under electromigration conditions. In section~3 we test the fidelity of the theory by showing the results of a  Monte-Carlo simulation of the temporal evolution step correlation function. Concluding remarks are contained in section~4.

\section{Theory}

In order to describe the dynamics of a step-edge evolving under electromigration conditions we employ   the usual Langevin formalism where each degree of freedom diffuses towards lower energy with a velocity that is proportional to the energy gradient, subject to random thermal fluctuations. The position of step edge is described by it's edge profile $y(x,t)$ where the x-axis is oriented parallel to the step edge. $y(x,t)$ is the position of the step edge at $x$ and at time $t$.

In this paper we consider the limiting case where the step motion occurs most easily by adatom diffusion along the step edge itself. Adatom exchange between the step-edge and the adjacent terraces, {\em via} attachment and detachment,  is neglected. It is well known that atomic diffusion along a step edge is driven by the step curvature~\cite{Mullins}, which generates a flux
\be
\bold{J}_{c} = \frac{D_s \delta_s \gamma}{k T}\nabla_s \kappa
\ee
where $\bold{J}_{c}$ is the curvature-driven flux, $D_s=a_{||}^2/\tau_h$ is the surface diffusion constant, $a_{||}$ is the atomic diameter, $\delta_{s}=a_\perp$ the width of an atom perpendicular to the step,  $\tau_h$ is the average time between hopping events and  $\gamma$ is the step stiffness. $\nabla _{s}\kappa$ is the gradient of the curvature along the step edge. Mass conservation determines the normal velocity of the step edge,  
\be
\frac{\partial y}{\partial t}\hat{n} = - \Omega \nabla_s \cdot \bold{J}_{c}
\ee 
where $\Omega$ is the area of a single atom and $\hat{n}$ is normal to the step edge.

The inclusion of a thermal noise term, $\eta$ and linearization of the above equation leads to the well-known equation of motion for a step edge~\cite{Bartelt92}
\be
\left(\frac{\partial }{\partial t} -\frac{\Gamma_{h} \gamma}{kT} \frac{\partial^4 }{\partial x^4}
\right) y(x,t) = \eta(x,t). \label{t_one}
\ee
where we have defined
\be\label{GAMMA}
\Gamma_{h}  = \frac{  a_{||}^{3}  a_{\perp}^{2} }{\tau_{h}}.
\ee

We model the effects of electromigration by adding to this equation of motion a term generated by a constant force, $F=Z^{\ast}e\bold{E}$,  felt by atoms at the step edge, which arises from the application of a electric field, $\bold{E}$, to the material oriented parallel to the y-axis.  $Z^{\ast}$ is the effective valence of an atom at the step edge.  The electromigration force generates an additional flux,
\be
\bold{J}_{EM} = \frac{D_s \delta_s Z^{\ast}e\bold{E_{||}}}{\Omega k T}= \frac{D_s \delta_s Z^{\ast} e\bold{E}}{\Omega k T}
 \left( 
\frac{y_x}{\sqrt{1+y_x^2}}
  \right)
\ee
where $E_{||}$ is component of the electric field along the step edge. $y_x$ indicates an x derivative.  

The energy of any step configuration, relative to the energy of the straight step, is determined by the step stiffness $\gamma$ and the electromigration force, $F$, felt by atoms at the step edge. If the force acts perpendicular to the step edge (i.e. when $F>0$, the force acts in the +y, or step down, direction) then, for small fluctuations, one can linearize the stochastic equation of motion for the step edge to obtain the following Langevin equation for the step dynamics,

\be
\left(\frac{\partial }{\partial t} -\frac{\Gamma_{h} \gamma}{kT} \frac{\partial^4 }{\partial x^4} 
- \frac{\Gamma_{h}  F }{kT a_{||}a_{\perp} } \frac{\partial^2 }{\partial x^2}
\right) y(x,t) = \eta(x,t) \label{one}
\ee
where $a_{||}$ and $a_{\perp}$ are the lattice parameter parallel and perpendicular to the step edge. The  noise term, $\eta$, describes the thermal fluctuations and is correlated because, in our model, of the random hopping of atoms occurs only between between adjacent step edge sites.

\be
\langle \eta(x,t)\eta(x',t') \rangle = 2 \Gamma_{h} \delta(t-t') \frac{\partial^2}{\partial x^2} \delta(x-x') \label{stat}
\ee
To first order, the electromigration force does not alter the correlation properties of the noise so that in the single-jump regime equation~\ref{GAMMA} remians valid.

Before proceeding, for notational simplicity we rewrite eqn.~\ref{one} as

\be
\left(\frac{\partial }{\partial t} - \alpha \frac{\partial^4 }{\partial x^4} 
\mp \alpha q_{c}^2 \frac{\partial^2 }{\partial x^2}
\right) y(x,t) = \eta(x,t) \label{four}
\ee
where

\be
\alpha \equiv \frac{\Gamma_{h} \gamma}{kT}, \hspace{2cm} 
q_{c} \equiv  \frac{1}{\sqrt{a_{||}a_{\perp}}} \sqrt{ \frac{ \left| F \right| }{ \gamma  }} \label{five}
\ee
The parameter $q_{c}$ depends only on the magnitude of the electromigration force and in eqn.~\ref{one} the $\mp$ denotes the force acting in the step down(-) or step up (+) direction.

In the case where there is no random thermal noise (i.e. $T \rightarrow 0$), eqn.~\ref{one} predicts that the amplitude of a small sinusoidal fluctuation of the step-edge profile, with wavevector $q$, evolves according to the following dispersion relation,

\be
i \omega = \alpha q^2 \left( q^2 \pm q_{c}^{2} \right)
\ee
This is the same dispersion relation as the one that describes step flow under growth conditions~\cite{Bales90,Pierre99,Ramana99}.  If the electromigration force acts in the step up (+) direction the step fluctuations are linearly stable. For a force acting in the step-down direction there exists a linear instability which initiates the well-known phenomenon of electromigration-induced step-wandering for fluctuations with wavevector smaller than $q_{c}$~\cite{Krug04,Degawa01,Zhao04}. The critical wavevector is an important parameter in our Langevin model that, from eqn.~\ref{five}, is determined by the ratio of the electromigration force and the step stiffness.

In order to determine the statistical properties of the solutions of eqn.~\ref{one} we take the usual approach and first derive the Green's function for the problem using standard Fourier transform methods,

\be
G(x, t | x', t') = 
\frac{i}{2\pi} \int_{-\infty}^{+\infty}  e^{-\alpha (k^4 + q_{c}^2 k^2 )(t-t')} e^{ik(x-x')} dk \label{green}
\ee
In terms of the Green's function, the displacement of the step at time t is:

\be
y(x,t'') = y(x,t') + \int_{-\infty}^{+\infty} \int_{t}^{t''}  G(x, t'' | x', t) \eta(x',t') dx' dt \label{corrdef}
\ee
We can now compute the time correlation function of the step edge, $G(t)$, after time $t=t''-t'$ has elapsed,

\be
g(t) \equiv<(y(x,t'')-y(x,t'))^2>
\ee
Substituting eqn.~\ref{green} into~\ref{corrdef} and using the correlation properties of the noise (eqn.~\ref{stat}) we obtain we find, after some calculation,

\be
g_{\pm}(t) =  \frac{\Gamma_{h}}{2^{1/4}\pi \alpha}  
\int_{-\infty}^{+\infty}  \frac{1}{( k^2 \pm q_{c}^2) }
\left ( 1- e^{-2 \alpha (k^4 \mp q_{c}^2 k^{2}) t}   \right) dk 
\ee
or, using substitution,

\begin{eqnarray}
g_{\pm}(t) & = & t^{1/4} \left( \frac{ \left( kT \right)^3 \Gamma_{h} }{ \gamma^3 } \right)^{1/4}  \frac{2^{\frac{3}{4}}}{\pi} \times \\
& &
\int_{0}^{+\infty}  \frac{1}{ (u^2  \pm 
\sqrt{ \alpha t
q_{c}^4 } ) }
\left ( 1- e^{-2u^4} e^{\mp 2 u^{2} \sqrt{\alpha q_{c}^4}  t  }   \right) du \nonumber \label{bart}
\end{eqnarray}

When there is no electromigration force (i.e. $|F| = 0$, $q_{c} = 0$), the integral in eqn.~\ref{bart} is clearly time-independent and we recover the result derived by Bartelt {\em et al.}~\cite{Bartelt92} where the $g(t)$ the step edge fluctuations evolves with the well-known $t^{1/4}$ scaling law characteristic of step-edge limited diffusion.

\be
\lim_{F \rightarrow 0} g_{\pm}(t) 
\equiv
 g_{0}(t) 
=
t^{1/4} \left( \frac{  \left( kT \right)^3 \Gamma_{h} }{ \gamma^3} \right)^{1/4} 
\frac{2 \ }{\pi} \Gamma \left(\frac{3}{4}\right)  \label{bart0}
\ee
It is helpful to re-express eqn.~\ref{bart0} in terms of the average time, $\tau_{0}$, that it takes for the mean-squared width of an initially straight step to reach a value equal to  $g_{0}^{2}(\tau_{0})=a_\perp^2$ (i.e. one square lattice spacing):

\be
 g_{0}(t) 
= a_{\perp}^{2}
\left( \frac{t}{\tau_{0}}
\right)^{1/4}
 \label{tau0}
\ee
where

\be
\tau_{0}
\equiv
 \tau_{h} 
  \left( \frac{ \gamma a_{\perp}^{2}}{ 2 kT a_{||}} \right)^{3} 
\left(  \frac{ \pi }{ 2^{\frac{1}{4}}\Gamma \left(3/4 \right)  } \right)^{4} \label{tau0def}
\ee

When the electromigration force is finite ( i.e. $|F| >  0$), it is apparent from eqn.~\ref{bart} that this scaling behavior is modified by the appearance of an explicit time dependence in the integral. This can be seen more clearly by defining a critical time, $\tau$, and a rescaled time, $\zeta=t/\tau$, where, 

\be
\tau = \frac{1}{\alpha q_{c}^4} =
2 \tau_{h}
 \left( \frac{kT}{Fa_{\perp}} \right)^{2}
\left( \frac{ \gamma a_{\perp}^2 }{2kT a_{||}} \right), \label{taudef}
\ee
Then, eqn.~\ref{bart} can be rewritten as

\be
g_{\pm}(t) =  a_{\perp}^{2}
\left( \frac{t}{\tau_{0}}
\right)^{1/4} I_{\pm} \left( \frac{t}{\tau}   \right)\label{14}
\ee
where
\begin{eqnarray}
 I_{\pm} \left( \zeta   \right) & \equiv & 
\frac{1}{ 2^{1/4} \Gamma(3/4)}  \times \\
& &
\int_{0}^{+\infty}  
\frac{1}{ (u^2  \pm \sqrt{ \zeta} )}
\left( 1
- e^{-2u^4} 
e^{\mp 2 u^{2} \sqrt{\zeta}}
   \right) du \nonumber  \label{15}
\end{eqnarray}
 $I_{\pm}$ is a universal function of the rescaled time, $\zeta$ and is normalized such that $I_{\pm}(t \rightarrow 0)=1$. The integral appearing in eqn.~\ref{15} is easily evaluated numerically and is shown in fig.~1 in which we display $I_{\pm}$  plotted as a function of the rescaled time $\zeta$ (i.e. time is expressed in units of $\tau$). The solid curves show $I_{\pm}$ obtained  for $F>0$ (step up) , $F<0$ (step down) electromigration forces (For $F=0$, $I_{\pm}(\zeta)=1$). 
 
  \begin{figure}
  \includegraphics[width=90mm]{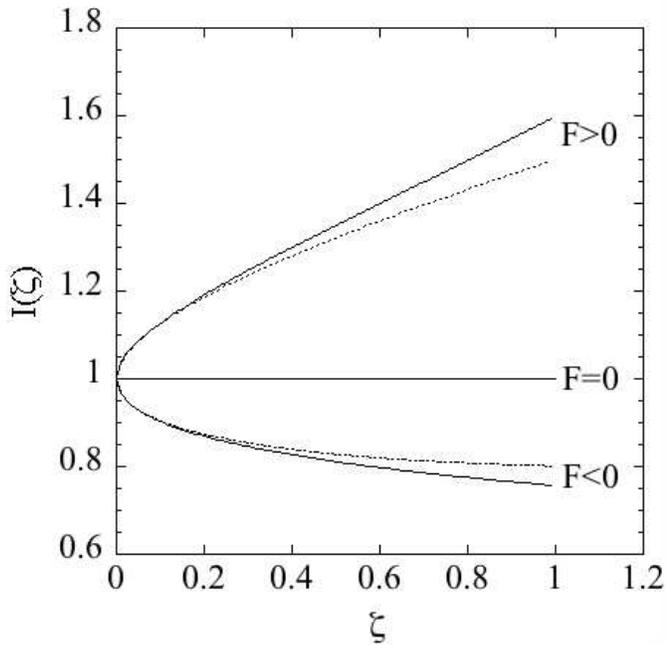}
 \caption{ \label{figure1}
 The integral function $I(\zeta)$ (eqn.~\ref{15}) plotted as a function of the rescaled time $\zeta=t/\tau$ for no electromigration force ($F=0$), the electromigration force in the step down direction ($F>0$) and in the step-up direction ($F<0$).
  }
 \end{figure}

 From fig.~1. it is apparent that $g_{\pm}(t)$ (eqn.~\ref{14}) deviates very significantly from the $t^{1/4}$ scaling behavior of $g_{0}(t)$ (eqn.~\ref{tau0}) as $t$ approaches $\tau$ ($\zeta \rightarrow 1$). These deviations begin to appear at earlier times, $t \sim 0.1\tau$, when the effect of the force on the evolution of the step fluctuations begins to be felt. This can be seen more clearly in fig.~2 which displays the correlation function of a step plotted as a function of time for $\tau_{0}=5s$ and $\tau=10^{+4}s$. The values for these parameters were chosen so that $\tau/\tau_{0} \sim 10^{4}$, a ratio typical of accelerated electromigration experiments. As noted above, deviations from $t^{1/4}$ scaling start to appear when $t \sim 0.1\tau =100s$. The results shown in figs.~1 and~2 have a simple qualitative interpretation; the short time behavior of the step fluctuations ($t  \ll \tau$) is completely dominated by the thermal fluctuations and the effect of the electromigration bias emerges only at later times. Such behavior is typical of  the dynamics of diffusion driven by weak external forces.
 
 \begin{figure}
 \includegraphics[width=90mm]{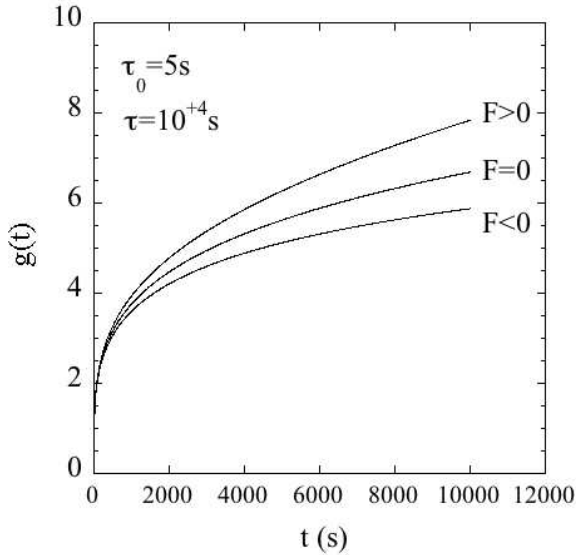}
 \caption{ \label{Finalfigure2}
The time corrleation function, $g(t)$, of the step-edge distribution predicted by the continuum Langevin model (eqns.~\ref{14} and~\ref{15}), plotted as a function of time for $\tau_{0}=5s$ and $\tau=10^{4}s$.
  }
 \end{figure}

It is instructive to perform a power series expansion  of the integral about $\zeta=0$ such that eqn.~\ref{14} becomes,

\be
g_{\pm}(t) = a^{2}
\left( \frac{t}{\tau_{0}}
\right)^{1/4}
  \left( 1\mp  a_{\frac{1}{2}} \left[ \frac{t}{\tau} \right]^{\frac{1}{2}} + 
a_{1} \left[\frac{t}{\tau} \right] 
+ O  \left[ \frac{t}{\tau} \right]^{\frac{3}{2}}
\ldots
\right) \label{series}
\ee
The expansion coefficents can be obtained analytically, 
\be
a_{\frac{1}{2}} =\frac{\pi}{6 \Gamma(\frac{3}{4})^2} = 0.3487, \hspace{0.75cm}
a_{1} =\frac{3}{20} = 0.1500, \hspace{0.75cm}
\ee
Shown as the dashed lines in fig.~1 are the results of this series approximation for  $I_{\pm}(\zeta)$ (eqn.~\ref{series}), evaluated up to, and including the terms linear in time. Clearly, this truncated expression is a reasonable approximation for $t \le 0.4 \tau$. 
 
\section{Simulation} 
 
In order to test the predictions of the continuum Langevin model described above, we developed a  Monte Carlo simulation of step edge fluctuations in the presence of an external force. In this model, atomic diffusion was restricted to the step edges with atoms jumping between adjacent step sites only. Only nearest-neighbor interactions on a square lattice were permitted ($a_{\perp}=a_{||}=a=1$) and were modelled by a single bond energy $\epsilon$. In order to describe the electromigration force, the atomic jump probabilities for motion parallel and anti-parallel to the force were biased by a small energy differential $\Delta U$. In terms of the electromigration force, $F$, and the lattice spacing perpendicular to the step edge, $a$; $\Delta U = \pm F a $.

Simulations were performed for steps of length $\ell = 10000a_{||}$ fluctuation on a square lattice. Periodic boundary conditions were employed. The bond energy was set to $\epsilon = 2.0 k_{B}T$ and the magnitude of the electromigration bias was  $|\Delta U | = 0.01 k_{B}T$, a factor of $10^{3}$ smaller than the typical binding energy of an atom to the step edge. This value was chosen to generate statistically significant deviations from the (no-force) $t^{1/4}$ scaling within reasonable simulation times. If $\epsilon = 0.1eV$ and it is assumed that $a \sim 1.5\AA$ (typical of metals) then this bias corresponds to an electric field with a magnitude of order $1000 Vcm^{-1}$ acting on an atom with effective valence of $Z^{\ast} \sim \pm 10e$. In actual accelerated electromigration experiments, a field of $0.1-1Vcm^{-1}$ is typical.

 \begin{figure}
 \includegraphics[width=90mm]{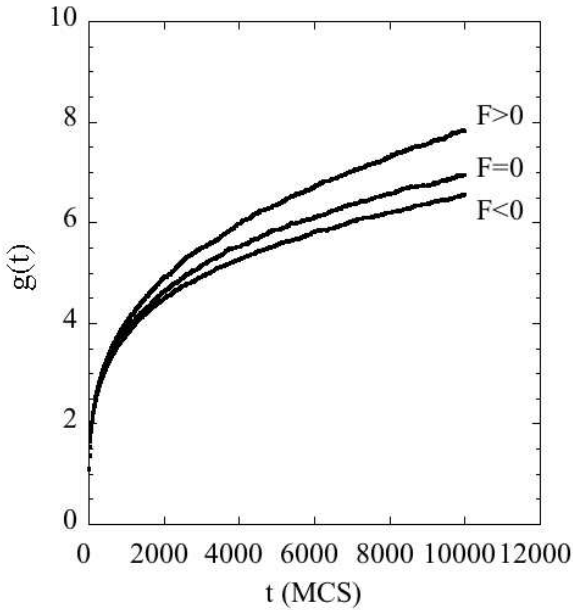}
 \caption{ \label{figure3}
  Results of the a kinetic Monte Carlo simulation where the correlation function of an isolated step is plotted as a function of the time (measured in Monte-Carlo steps per step-edge site, the lattice spacing is $a=1$). Shown is $g(t)$ obtained when the electromigration force acts in the step-up and step-down directions and  when $\Delta U = 0$ (i.e. no electromigration force is acting at the step edge). The curves shown were obtained by averaging the data from 200 randomly generated replicas after each was equilibrated for $10^{5}$ Monte Carlo steps.   
}
 \end{figure}

Figure~3 shows the results of the simulation where the correlation function of an isolated step is plotted as a function of the time measured in Monte-Carlo steps per step-edge site (MCS). Shown is $g(t)$ obtained when the electromigration force acts in the step-up and step-down directions and  when $\Delta U = 0$ (i.e. no electromigration force is acting at the step edge).We define one Monte Carlo step to be equal to the average time needed for every atom at the step edge to attempt a hop. The results shown in fig.~3 were obtained by averaging the data from 100 randomly generated replicas after each was equilibrated for $10^{5}$ Monte Carlo iterations per site.    Comparing the simulation results (fig.~3.) to the predictions of the Langevin theory (fig~4.) it is apparent that the qualitative features of the continuum theory are reproduced by the Monte Carlo simulation. These same data are presented in the form of a log-log plot in fig.~4. In the absence of the force ($F=0$ , $\Delta U = 0$), a least-squares fit of the no-force simulation data shows that $g_{0}(t)$ is very well fit by a $t^{\beta}$ power-law where $\beta=0.25\pm 0.01$. Therefore, when there is no electromigration force present in the simulation, the correlation function of the step evolves according well-known $t^{1/4}$ scaling, as  predicted by the Langevin analysis (eqn.~\ref{tau0}). By least squares fitting the simulation results to eqn.~\ref{tau0} we obtain a value of $\tau_{0}=5$ MCS. 

 \begin{figure}
 \includegraphics[width=90mm]{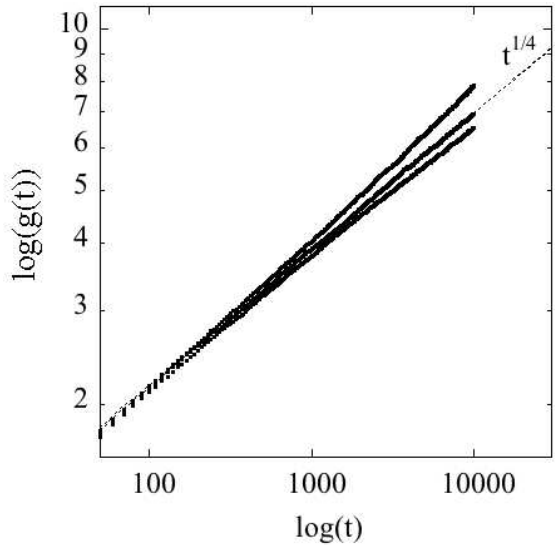}
 \caption{ \label{figure4}
 A log-log plot of the simulation data shown in fig.~3. The dashed line shows the best fit of a power law, $(t/\tau_{0})^{\beta}$, to the no-force ($F=0$) data;  $\beta=0.25$, $\tau_{0}=5$ MCS. 
}
 \end{figure}

We now consider the results of the simulation obtained for a finite electromigration force, also shown in figs.~3 and 4. Equation~\ref{series} suggests that the value of $\tau$ can be extracted from the simulation results by considering the scaling of the difference between the correlation functions for the force in the up and down step directions:

\be
\Delta (t) = \frac{g_{+} - g_{-}}{g_{0}} = 2 a_{\frac{1}{2}} \left[ \frac{t}{\tau} \right]^{\frac{1}{2}} +{\cal O}\left[ \frac{t}{\tau} \right]^{\frac{3}{2}} \ldots \label{diff}
\ee
For the simulation results, this normalized difference is plotted in fig.~5. The behavior of this quantity is well fit by the leading order term in eqn.~\ref{diff} from which we obtain a value of $\tau=62000 \pm 10000$ MCS.

 \begin{figure}
 \includegraphics[width=90mm]{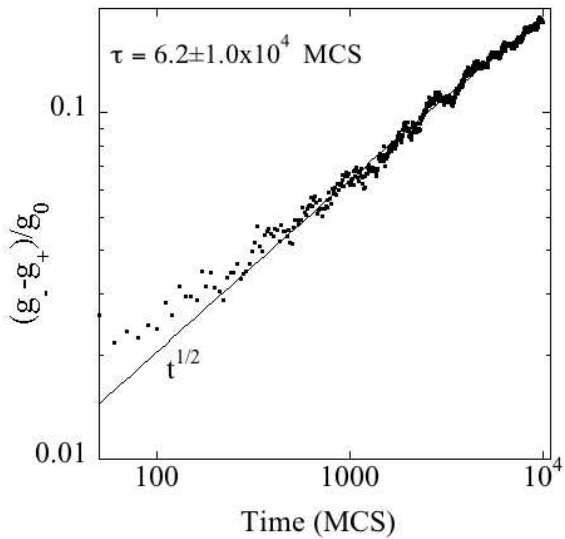}
 \caption{ \label{figure5}
For the simulation results shown in figs.~3 and~4; a log-log plot of the normalized difference (eqn.~\ref{diff}) plotted as a function of time (MCS). The behavior of this quantity is well fit by the leading order term in eqn.~\ref{diff} ($t^{1/2}$, dashed curve) from which we obtain a value of $\tau=62000 \pm 10000$ MCS.  
}
 \end{figure}

For comparison with the fits to the continuum Langevin model , we can estimate $\tau$ from the microscopic parameters employed in the discrete Monte Carlo model used to generate the simulation data. Combining eqn.~\ref{taudef} with the step stiffness appropriate for our model 

\be
\frac{\gamma a_{\perp}^{2}}{2 k T a_{||}} = \frac{\epsilon a_\perp}{2k_B T a_{||}}
\ee
we obtain an estimate of $\tau$ in units of MCS:

\be
\tau =  2  \left(\frac{\tau_{h}}{\tau_{a}} \right) \left(  \frac{k_B T}{\Delta U} \right)^{2} \left( \frac{\epsilon a_\perp}{2 k_{B} T a_{||}}\right) 
\label{taurat}
\ee
The ratio of the hopping time in the Langevin model , $\tau_{h}$ to the time between hopping attempts in the Monte Carlo simulation $\tau_{a}$ can be obtained by monitoring the success rate of hops between adjacent lattice sites in the simulation. We find that $\tau_{a}/\tau_{h} = 3.6\pm0.1$. Thus our estimate for the value of $\tau$ in the Monte Carlo simulations, used to generate the data shown in figs 3-5, is $\tau = 62000 \pm 10000$ MCS. Clearly, the  agreement between the continuum Langevin theory ($\tau = 71000$ MCS ) and the microscopic model ($\tau = 62000 \pm 10000$ MCS) is  reasonable.
	Finally we note that in the high temperature limit ($k_{B}T \gg \epsilon$) the ratio of the electromigration bias to the binding energy at a step edge, $\epsilon$ is related directly to $\tau$ that would be obtained from experiment:	
\be
\frac{\tau}{\tau_h} \approx \frac{1}{2} \left(  \frac{\epsilon}{\Delta u}   \right)^{2}
\ee	
where $\tau_{h}$ can be determined from eqn.~\ref{tau0def}, if the step stiffness is known.

\section{Conclusions}

The temporal evolution of step-edge fluctuations under electromigration conditions has been analyzed using a continuum Langevin model for the case where diffusion is limited by mass transport along the step edge. We find that the presence of the electromigration force, felt by atoms at the step edge, causes deviations from the usual $t^{1/4}$ scaling of the terrace-width distribution driven by thermal fluctuations alone.  We have identified  a critical time $\tau$ that is a function of the three energy scales in the problem: $k_{B}T$, the step stiffness, $\gamma$, and the bias associated with adatom hopping under the influence of an electromigration force, $\pm \Delta U$. For ($t  <  \tau$), the correlation function  evolves, to a good approximation, as a superposition of $t^{1/4}$ and $t^{3/4}$ power laws. For all $\tau$ a closed form expression was derived. This behavior was confirmed by a Monte-Carlo simulation using a discrete model of the step dynamics. Finally we propose that the magnitude of the electromigration force acting upon an atom at a step-edge could be determined directly by careful measurement and analysis of the statistical properties of step-edge fluctuations on the appropriate time-scale.

\begin{acknowledgments}
We acknowledge helpful discussions with E.D. Williams.

This work has been supported by the US Department of Energy grant DE-FG02-01ER45939 and by the NSF-Materials Research Science and Engineering Center under grant DMR-00-80008.
\end{acknowledgments}

\bibliography{langevin}

\end{document}